\documentstyle[11pt,aaspp4]{article}

\begin{document}

\title{Galaxy Morphologies in the Cluster CL1358+62 at
z=0.33\footnote{Based on observations with the NASA/ESA Hubble Space
Telescope and the W.~M.~Keck Observatory}}

\author{Daniel Fabricant}
\affil{Center for Astrophysics, 60 Garden St., Cambridge, MA 02138\\}
\and
\author{Marijn Franx and Pieter van Dokkum\footnote{Present address:
California Institute of Technology, MS 105-24, Pasadena, CA 91125}}
\affil{Leiden University, P.O.~Box 9513, NL-2300, Leiden, the Netherlands\\}
\centerline{\it Accepted for publication in the Astrophysical Journal}
\vspace{.2in}

\begin{abstract}

We describe the morphological composition of a sample of 518 galaxies
in the field of CL1358+62 at z=0.33, drawn from a large {\it HST}
mosaic covering 53 sq.~arcmin.  The sample is complete to $I$=22,
corresponding to $M_V$=-18.5 in the rest frame.  The galaxy
morphologies have been independently classified by the authors of this
paper and by Alan Dressler.  Dressler's classifications place our work
in context with the previous MORPHS study, and allow us to estimate
the scatter between different sets of visual classifications.

We restrict most of our analysis to the brighter part of the sample,
$I<21$ ($M_V<-19.5$), where the scatter between the two sets of
classifications is $\sim$1 in morphological type.  The scatter doubles
at $I=22$, presumably due to the lower signal-to-noise and poorer
sampling of faint, small galaxy images.  To $I$=21 the two sets of
classifiers agree on the fraction of early type galaxies
(elliptical+S0): 72\%.  We conclude that CL1358+62 does not contain
the large population of spiral galaxies found in other studies of
clusters at $z\sim0.3$, and that there is probably a significant
spread in the degree of cluster evolution at intermediate redshift.

The two groups of classifiers differ on the relative fraction of S0
and elliptical galaxies.  We show that the distributions of
ellipticities and bulge/total light cannot resolve this discrepancy.
Nonetheless, we can derive significant constraints on physical models
for the evolution of the galaxy population in CL1358+62.  The higher
ratio of S0 to elliptical galaxies (1.6) found by DF/MF/PvD requires
that the evolution preserve the relative fraction of elliptical, S0
and spiral galaxies.  Alternately, the lower ratio (1.1) found by AD
requires that the evolution preserve the early-type to spiral ratio
while increasing the S0 to elliptical ratio.  In the latter case, a
possible evolutionary mechanism is accretion of galaxies that
predominantly evolve to S0's between $z$=0.33 and the present.

We use our large body of spectra to make the correspondence between
spectral and morphological type.  Our data follow the pattern seen in
the field at low redshift: emission line spectra are more prevalent
among the later morphological types.  The 11 identified k+a galaxies
(absorption line spectra with strong Balmer lines) have S0--Sb
morphologies.

\end{abstract}

\section{Introduction}

The WFPC2 camera on the Hubble Space Telescope ({\it HST}) has made it
possible to determine the morphologies of galaxies at intermediate
redshift and beyond.  It has been known for some time that the
photometric and spectral properties of galaxies in intermediate
redshift clusters differ from galaxies in nearby clusters; the
population of blue, star-forming galaxies and post-starburst galaxies
is larger at intermediate redshift, see e.g., \cite{bo84}, \cite{dr87},
\cite{gd88}, \cite{cs87}, and \cite{dr92}.  Given the correlation between the
spectral and morphological properties of galaxies (\cite{mm57},
\cite{mo69}, \cite{ke92}), we might be able to detect a corresponding
evolution of galaxy morphology in intermediate redshift clusters.
However, the evolution of morphology is likely to be more subtle than
spectral evolution, since galaxies of the same morphological type can have
significantly different star formation rates (e.g. \cite{ja99}).

In the most ambitious study of this sort with WFPC2 to date, the
MORPHS group have classified over 1200 galaxies in 10 clusters at
0.37$<$z$<$0.56 (\cite{sm97}).  They find that S0 galaxies are less
common than in low redshift clusters and that the ratio of S0's to E's
within a radius of $\sim$600 kpc (for H$_0$=50, q$_0$=0.5) decreases
with redshift, falling from 2 in low redshift clusters to less than
0.5 at $z=0.5$.

\cite{an97} (see also \cite{an98}) have independently classified
galaxies in a WFPC2 image of Cl0939+4713, one of the less concentrated
clusters in the MORPHS sample.  They find a ratio of S0's to E's of
$\sim$2, quite comparable to a low-$z$ reference sample in the Coma
Cluster.  However, they classify 40-50\% of the galaxies in
Cl0939+4713 as spirals (S), in contrast with 20-30\% S in a comparable
region of the Coma Cluster.  The sample of galaxies in Cl0939+4713 is
relatively small ($\sim$70), and redshifts are available for less than
one third of these.

\cite{co98} present a study of the morphological types in three
clusters at $z$=0.31, also using WFPC2 images.  There is substantial
overlap between the authors of this paper and the MORPHS group, and
the two groups have attempted to adopt a consistent morphological
system.  At $z$=0.31, Couch et al.~find an excess of S's, with an
abundance at small radii ($\sim$400 kpc for $H_0$=50, $q_0$=0.5)
approximately twice that in low-$z$ reference clusters.  However,
averaged over the three clusters, within 400 kpc, they find a ratio of
S0's to E's at most slightly depressed relative to regions of
comparable galaxy surface density in low-$z$ clusters.  \footnote{It is
important to account for the morphology-density relation when we
consider the morphological content of clusters.  At high galaxy
densities, typically found in the cores of clusters, the low-$z$
reference population becomes increasingly dominated by E's,
\cite{dr97}}.  Note, however, that the average $z$ of the MORPHS
clusters is larger (0.46).

\cite{lu98} report the morphological types in two more distant
clusters at $z\sim$0.9.  One cluster, CL0023+04, appears to be
composed of two low velocity dispersion groups, and contains
predominantly S's. The other, CL1604+43, with a velocity dispersion of
$\sim$1200 km s$^{-1}$, contains $\sim$76\% early-types.  In the
latter case, the S0/E ratio is found to be 1.7$\pm$0.9.  This result
is sensitive to the assumed morphological composition of the
foreground/background population, but is evidence that the
S0/E ratio does not decline smoothly with $z$.

Our approach is complementary to the \cite{dr97}, \cite{an97}, and
\cite{co98} studies which predominantly describe the galaxy
morphologies in cluster cores. We use mosaics of WFPC2 fields to study
the galaxy population in a larger region (allowing larger galaxy
samples per cluster), and we have acquired large numbers of spectra of
cluster galaxies.  The spectra remove ambiguity about cluster
membership and allow us to directly connect the morphological and
spectral properties of the galaxies.  Our sample of clusters is x-ray
selected, with x-ray luminosities exceeding 4$\times$10$^{44}$ erg
s$^{-1}$ in the 0.2--4.5 keV band.  

In CL1358+62 at $z$=0.33, we have drawn a complete sample of 518
galaxies to a magnitude limit $I$=22 from a WFPC2 mosaic image of
CL1358+62 covering 53 square arcminutes.  Spectra for 276 of the 518
galaxies in the morphological sample were previously obtained at the
Multiple Mirror and William Herschel Telescopes.  The color-magnitude
relation of the 194 spectroscopically confirmed cluster members in the
{\it HST} mosaic (3 are fainter than $I$=22) has been previously
described in \cite{vd98}.  The spectroscopic properties of 232 cluster
members (some outside the {\it HST} mosaic), as well as the cluster
dynamics have been described in \cite{fi98}.

Our objectives in this paper are fourfold.  (1) We introduce the
morphological classification techniques that we will apply to our entire
sample of clusters.  (2) We classify the galaxies in CL1358+62 at
z=0.33, comparing our classifications with those of an experienced
external researcher, Alan Dressler.  Our deep sample with two
independent classifications provides a useful assessment of the
scatter between WFPC2 visual morphological classifications at
intermediate redshifts.  (3) We describe the robust,
classifier-independent conclusions and explore the physical
implications of the differences between the two sets of
classifications.  (4) We connect the spectral and morphological types
of the cluster galaxies.

The paper is organized in the following fashion.  In $\S$ 2, we
describe the photometric catalog from which the morphological sample
was drawn.  The two set of morphological classifications are discussed
in $\S$ 3. The morphological composition of the cluster and evidence
for morphological evolution are presented in $\S$ 4.  The connection
between the morphological and spectral properties of the galaxies is
made in $\S$ 5.  $\S$ 6 contains a brief discussion and conclusions.

\section{Photometric Catalog}

Our photometric catalog, from which we draw galaxies for morphological
classification, is derived from the {\it HST} F814W mosaic image.  The
techniques used to construct the mosaic are described in \cite{vd98}.
Our use of the mosaic image, instead of individual WFPC2 CCD frames,
slightly compromises the accuracy of the photometry, but considerably
simplifies the source detection problem by eliminating most of the
boundaries.  Because our goal is to select a sample of galaxies for
morphological classification to a consistent magnitude limit, rather
than to perform precision photometry, this is a beneficial tradeoff.
We used the SExtractor package, described by \cite{se96}, to detect the
galaxies and perform the photometry.  We use a detection and analysis
threshold of 24.5 mag arcsec$^{-2}$, and a zeropoint of 30.546 (3600 s
exposures, May 1997 WFPC2 SYNPHOT update) to convert from instrumental
magnitudes to a Cousins $I$ magnitude\footnote{ Throughout the paper
we consistently refer to a Cousins $I$ magnitude, which is very close
to the natural F814W system referred to a Vega zeropoint
(\cite{ho95})}.  The adopted zeropoint is the average of the value for
the 3 WFPC2 CCDs, which vary by $\sim$$\pm$0.01 magnitudes.  We use
the SExtractor total magnitude estimator (see \cite{se96}), which is
insensitive to the analysis threshold.

In order to allow convenient comparisons with the results of the
MORPHS group (\cite{dr97}), we calculate a conversion of observed $I$
magnitudes (for $z<$0.6) to a $V$ magnitude in the rest frame.  Some
of the MORPHS classifications used WFPC2 F702W filter data, so we also
derive a consistent conversion from F702W magnitudes to a rest frame
$V$.  The intent in \cite{dr97} was to work to a consistent limit of
$M_V$=-20, but a transcription error from Table 9 of \cite{ho95} led
to the adoption of a deeper limit of $M_V$$\sim$-19. (Here and
throughout we use $M_V$ to refer to a rest frame $V$ absolute
magnitude.) In contrast with \cite{dr97}, we also apply an
evolutionary correction to allow a closer comparison with low $z$
clusters.  We adopt the \cite{dr97} cosmological model (H$_0$=50 km
s$^{-1}$ Mpc$^{-1}$, q$_0$=0.5)

We use four numbers in addition to the distance modulus to convert
from the WFPC2 magnitudes to $M_V$: (1) the conversion from the
Vega-referenced WFPC2 natural filter system to Cousins filter bands,
(2) the $z$-dependent $K$ correction to transform the observed Cousins
$R$ and $I$ magnitudes to the rest frame, (3) the estimated rest frame
galaxy colors, and (4) a $z$-dependent evolutionary correction.  Each
of these numbers depends on the spectral energy distribution of the
galaxies, so the accuracy of this procedure is limited.  We use these
numbers, however, only to choose sample limiting magnitudes
appropriately scaled with $z$.  We adopt the approximate expressions
described below for these conversions.

We first fit a linear relation to the synthetic transformation of
F702W to $R$ as a function of $V-R$ (valid for $V-R$$<$1.5), using
results from Fig.~10 of \cite{ho95}:

$R-F702W=0.31(V-R)$~~~~~(1a)

\noindent We take the $V-R$ colors of galaxies as a function of $z$
from \cite{fr94}, with a morphological mix of 70\% E and 30\% Sbc to
convert (1a) to:

$R-F702W=0.31(0.5373 - 0.9132z + 10.17z^2 - 11.21z^3)$~~~~~(1b)

\noindent To a good approximation (0.1 mag), Fig.~9 of \cite{ho95} shows:

$I-F814W=0$~~~~~(1c)

\noindent We fit polynomials to the average $K$ corrections of \cite{fr94} and
\cite{pog97}, using 70\% E and 30\% Sbc or Sc contributions:

$K_R=0.4293z + 3.807z^2 - 2.903z^3$~~~~~(2a)

$K_I=0.4910z + 0.4836z^2$~~~~~(2b)

\noindent From \cite{fr94} we take:

$V_{rest}-R_{rest}=0.53$~~~~~(3a)

$V_{rest}-I_{rest}=1.10$~~~~~(3b)

\noindent From the the early-type galaxy fundamental plane study of \cite{ke97}
we derive an evolutionary correction:

$EC_{V_{rest}}=-0.77z$~~~~~(4)

\noindent The corrections are applied in the following fashion:

$I  = M_V + DM + EC_{V_{rest}} - (V_{rest}-I_{rest}) + K_I$ 

\noindent Here, $DM$ is the distance modulus.  Applying these expressions to
CL1358+62, with a distance modulus at $z=0.3283$ of 41.62, we find
that the MORPHS limit of $M_V$=-20 corresponds to an observed
$I$=20.5.

Our photometric catalog completeness extends below $I=23$, but we have
limited our morphological sample to $I=22$ to increase the reliability
of the morphological classifications.  At $I=22$, we attain a
signal-to-noise ratio (S/N) that is very similar to the S/N attained
at the $I=23$ classification limit adopted by the MORPHS group for
their deeper images (\cite{sm97}).  The standard deviation of the sky
noise in our (3600 s) CL1358+62 image has an equivalent surface
brightness of 24.4 mag arcsec$^{-2}$, as compared with 25 mag
arcsec$^{-2}$ typical for the ($\sim$12600 s) MORPHS images.  A crude
scaling relation can be derived by assuming that all galaxies have the
same surface brightness, and that our criterion is that
classifications of equal reliability require the same number of pixels
at the same S/N.  For a shallower image, we can effectively bin the
image of a brighter galaxy that is $n$ times larger $n\times n$ pixels
to achieve a S/N that is $n$ times better.  To recover a sky S/N
deficit of 0.6 mag, we need to set a limit $\sim$1.2 mag brighter.

\section{Morphological Classification}

There are 518 galaxies in the morphological catalog after removing
stars and image artifacts.  In the rest frame, the filter central
wavelength corresponds to $\sim$6100 {\AA}.  We use the F814W images
for morphological classification because they are deeper than the
F606W images.  \cite{dr97} have used either F702W and F814W images for
classification at comparable redshifts (between 0.37 and 0.41).

Below, we describe two sets of morphological classifications.  The
first set was carried out by the authors of the paper using the
techniques described in $\S$3.1.  We realized that our classifications
would be of greater value if we could compare them with morphological
classifications from an external expert.  Alan Dressler (AD) kindly
agreed to independently classify the entire sample.  Dressler is a
member of the MORPHS group, and most importantly, has classified a
large reference sample of low-$z$ cluster galaxies.  The strength of
the AD classifications is that the same experienced classifier has
classified the galaxies at low and intermediate redshifts.
Consistency has obvious benefits when searching for evolution.

However, it is important to keep in mind the difficulty of classifying
these relatively faint and small galaxies from WFPC2 images.  The
brighter cluster ellipticals and S0's in CL1358+62 ($I\sim19$) have
effective radii (half-light radii), r$_e$, of 0.5$^{\prime\prime}$ to
0.6$^{\prime\prime}$.  The faintest galaxies in the sample ($I\sim$22)
have r$_e$$\sim$0.3$^{\prime\prime}$.  The 50\% encircled energy
radius of WFPC2 stellar images measured from our frames is about
0.17$^{\prime\prime}$, so the numbers of meaningful information
elements are small: typically 20 to 40.  For this reason, we cannot be
sure that agreement on a common morphological system is the most
significant issue.  Even the most experienced classifier may not
account for all the systematic differences between the low $z$
(photographic) and the intermediate $z$ WFPC2 images.  By comparing
the two sets of morphological classifications we will be able to
discern which aspects of the classification are most robust.

\subsection{DF/MF/PvD Classifications}

The 518 sample galaxies were independently classified along the
revised Hubble sequence by DF, MF and PvD from 96$\times$96 pixel
``postage-stamps'' drawn from both the mosaic and original individual
CCD images.  We referred frequently to \cite{sa61} and \cite{sa87}, as
well as to artificially redshifted digital images drawn from the
Nearby Field Galaxy Survey (\cite{ja99}).  We assigned numerical types
as follows: -5 (elliptical), -4 (elliptical or S0), -2 (S0), 0 (S0 or
Sa), 1 (Sa), 3 (Sb), 5 (Sc), 7 (Sd), 9 (Sm), 10 (Im) and 99 (peculiar
or merger).  Intermediate types (2, 4, 6, and 8) were also assigned.

For 80\% of the galaxies, the three independent classifications span a
range of three or fewer numerical types (we consider -4, -2 and 0 to be
adjacent numerical types), and agree exactly for 34\% of the sample.
For 6\% of the sample, the object is considered unclassifiable by one
or more of the authors, or else the classifications disagree wildly.
In these cases, we assign a numerical type of 999.  The remaining 14\%
are classified type 1 or later by all the authors, but with a broader
range of classifications.  Table 1 lists the combination rules used to
assign types where the agreement is not exact, including type 15
for indeterminate late types.

\subsection{AD Classifications}

Following the completion of the DF/MF/PvD classifications, AD
independently classified the CL1358+62 galaxies according to
techniques described in \cite{sm97}.  AD does not assign galaxies to
the DF/MF/PvD intermediate types -4 (E or S0) or 0 (S0 or Sa) in the
same fashion, preferring to subdivide these into E/S0 (-4), S0/E (-3),
S0/Sa (-1) and Sa/S0 (0).  When discussing the cluster population in
broad terms, AD's types -5 and -4 are combined into E, types -3, -2 and -1
into S0, and etc.  DF/MF/PvD split the contents of the -4 bin equally
into E and S0, and the 0 bin equally into S0 and Sa.  When converting
AD's descriptive types into numerical types, we have placed a few
galaxies into the merger (99) bin where an individual catalog object
corresponds to an interacting or merging pair of objects.

Augustus Oemler, another member of the MORPHS group, independently
classified 307 galaxies from the CL1358+62 sample ($I$=22 limit) to check
if AD's classifications adhere to the MORPHS system.  Oemler's
classifications agree very well with AD's overall, with $\sim$10\%
more S0's by number, and a corresponding decrease in the numbers of
E's.  The number of spirals is identical in both classifications.

\subsection{Quantitative Comparison of the Classifications}

Like any measurement, classifications will suffer from random and
systematic errors.  We can estimate these errors by comparing multiple
sets of classifications. Ideally, such a comparison should be based on
classifications using independent imaging data.  Here, both sets of
classifiers (DF/MF/PvD and AD) worked from the same data set, so we
may underestimate the errors.  Nonetheless, this comparison is quite
interesting.  Figure 1 shows the difference between DF/MF/PvD and AD
morphological types for individual galaxies as a function of
magnitude.  As expected, the differences increase at fainter
magnitudes.  To avoid creating artifical gaps in Figure 1, we
condensed and shifted the numerical types of the early type galaxies
for this presentation: -3 for E's, -2 for E/S0's and S0/E's, -1 for
S0's, 0 for S0/Sa and Sa/S0. For types Sa and later, the ``normal''
types were used.  Galaxies which were classified as merger, peculiar,
or unclassified by one group were ignored.

Figure 2 is a scatter diagram comparing the two sets of
classifications.  The scatter is dominated by random differences, but
there is some evidence for systematic differences for the early type
galaxies. We return to this point below. The scatter between the
classifications has been measured by calculating the mean absolute
deviation, and normalizing it to the mean absolute deviation of a
gaussian with an rms of 1. This measurement of the scatter is much
more robust than the RMS of the differences.  The result is shown in
Figure 3.  The scatter is $\sim$1 for bright galaxies, but increases
strongly faintwards of $I=21$.  The appendix describes how classification
errors might systematically bias our population estimates;
unfortunately we cannot simply calculate correction factors.  In what
follows, we restrict most of our analysis to the subset of galaxies
with $I<21$. 

\subsection{Classification of Ellipticals and S0s}

Figure 2 shows a systematic difference in the DF/MF/PvD and AD
classifications of early type galaxies.  This is not surprising as the
division between E and SO is a difficult problem in visual
classifications.  We might hope that looking at two objectively
determined structural parameters for the galaxies, ellipticity and
the bulge/total light ratio, might be helpful in resolving this issue.
For example, \cite{sm97} compare the ellipticity distributions of the
MORPHS E and S0 intermediate-$z$ galaxies with those from an
\cite{an96} study of Coma Cluster galaxies as a consistency check of
the MORPHS classifications.

We have therefore examined the structural properties of the CL1358+62
galaxies measured by the Medium Deep Survey (MDS)
group\footnote{The MDS catalog is based on observations with the
NASA/ESA HST, obtained at the STSCI, operated by AURA.}, \cite{ra99}.
The MDS group has published structural properties for 70\% of the
CL1358+62 galaxies in our morphological catalog.  The missing 30\% are
located near frame edges or in one mosaic frame that is not yet
included in the MDS database.  The MDS group has chosen the best
fitting of four structural models for the objects in their catalogs:
star, disk, bulge, or disk + bulge, accounting for the point spread
function of HST.  If a galaxy model is chosen, the MDS group fit the
ellipticity of the disk, bulge or disk and bulge separately as
appropriate.  We focus on two structural parameters derived from these
fits: the ratio of bulge to total light, and the weighted ellipticity.
The ratio of bulge to total light follows trivially from the best fit
model, or from the fits to the disk + bulge model.  We calculate the
weighted ellipticity from the MDS disk and bulge ellipticities,
weighting by the fractions of light in the disk and bulge.

In Figures 4 and 5 we plot the distributions of the bulge to total
light ratio for the E and S0 galaxies with $I<21$ in the DF/MF/PvD and
AD samples, respectively.  Excepting the differences in numbers, the
distributions for both E's and S0's look remarkably similar for the
two sets of classifications.

In Figures 6 and 7 we plot the distribution of weighted ellipticities
for the same E and S0 galaxies ($I<21$), again showing the DF/MF/PvD
and AD samples independently.  Caution must be exercised when
comparing to these to other measures of ellipticity since the MDS
numbers are corrected for the HST PSF.  A larger fraction of the AD
E's have ellipticities exceeding 0.2, but the mean AD E ellipticity is
0.22, only slightly larger than the DF/MF/PvD mean of 0.17.  The
ellipticities of the two samples of S0's are similar: AD finds a mean
ellipticity of 0.46 and DF/MF/PvD find a mean ellipticity of 0.40.
The sizable difference in the numbers of E and S0 galaxies found by
the two sets of classifiers is not reflected in a large difference
between the structural parameters for the two samples of E's or S0's.
We conclude that although the total number of early-type galaxies is
well established, the SO to E ratio in CL1358+62 is uncertain.

\section{Morphological Composition of CL1358+62 and Evidence for Evolution}

We have learned that the differences between the two sets of
classifications rise steeply below $I=21$, suggesting this as a
practical limit for our morphological study.  This is also
conveniently close to the effective limit of our spectroscopic
completeness.  We have redshifts for 277\footnote{In one case, a
spectrum was obtained of a pair of galaxies separated by
0.7$^{\prime\prime}$, one (\#1295) at $I=21.17$ and the other (\#1297)
at $I=20.92$.  We have assigned the measured velocity (cz=99657) to
both galaxies. Galaxies \#1481 and \#1483 may have both fallen within
the spectrograph slit.  We assign the measured velocity (cz=99792) to
\#1483, which is 1.2 mag brighter.} of the 518 galaxies in the
morphological sample, and 191 of these are cluster members by the
criteria given in \cite{fi98}: 0.31461 $<$ z $<$ 0.34201.  The
spectroscopic completeness is 89\% for the galaxies brighter than
$I=20.5$ (corresponding to $M_V=-20)$, falling to 59\% for
$20.5<I<21$, 33\% for $21<I<21.5$, and 9\% for galaxies with
$21.5<I<22$.

Table 2 lists positions, $I$ magnitudes, both sets of morphological
classifications, and radial velocities for galaxies brighter than
$I$=21.

\subsection{Morphological Composition}

We may study the morphological composition of the subsample of known
members, which is nearly complete to the MORPHS's depth of $M_V=-20$,
without concern about background galaxy subtraction.  The subsample of
known cluster members to $I=20.5$ (or $M_V=-20$, see $\S$ 2) contains
138 galaxies.  DF/MF/PvD classify 27$\pm$4\% of these as E, 44$\pm$6\%
as S0, 29$\pm$5\% as S, with 1 unclassified galaxy. (The errors here
and in the following discussions account only for the Poisson
statistics of the number of galaxies per classification bin.)  AD
classifies 35$\pm$5\% as E, 38$\pm$5\% as S0, and 27$\pm$4\% as S.  In
both cases, the total early type population is $\sim$72\%.  To
$M_V$=-20, then, the only difference between the two groups of
classifiers is the relative numbers of E's and S0's.

We consider also the complete photometric sample to a depth of $I=21$
(or $\sim$$M_V=-19.5$), where a small background correction is
necessary.  To $I=21$ our sample contains 298 galaxies, for which we
have 236 redshifts (79.2\% completeness).  Of the 236 galaxies with
redshifts, 65$\pm$8 are nonmembers, yielding a foreground/background
count of 82$\pm$10 after correcting for the spectroscopic
completeness.  We determine the morphological composition of the
foreground/background galaxies directly from our spectroscopic sample,
which contains 86 foreground/background galaxies.  DF/MF/PvD classify
3\% E, 13\% S0, 77\% S, and 6\% mergers.  AD classifies 11\% E, 9\%
S0, 71\% S, and 9\% mergers.  We average these classifications, and
adopt a background composition of 7\% E, 11\% S0, 74\% S and 8\%
mergers.  For comparison, \cite{dr97} adopt a morphological
composition of 10\% E, 10\% S0, and 80\% S.

After correcting for background, to $I=21$, DF/MF/PvD find a
population of 25$\pm$4\% E's, 46$\pm$5\% S0's, 29$\pm$6\% S's, and
0$\pm$1\% mergers.  AD finds 36$\pm$4\% E's, 36$\pm$5\% S0's,
28$\pm$6\% S's, and 1$\pm$1\% mergers.  These results are indistinguishable
from those for the brighter spectrosopic sample, with $\sim$72\%
early-types in both sets of classifications.

\subsection{Morphological Evolution}

We search for morphological evolution in CL1358+62 by comparing its
population with that of equivalent low-$z$ clusters.  Judging which
low-$z$ clusters are equivalent is somewhat uncertain, but we take
as an approximation low-$z$ clusters with a similar number of galaxies
within a fixed metric aperture, allowing us to correct for the effects
of the morphology-density relation.  We use the nearby cluster catalog
of \cite{dr80}, reanalyzed and summarized in \cite{dr97} as a
benchmark.  We use the subset of high concentration clusters (10 of
55) in the low-$z$ sample for comparison, because CL 1358+62 was
selected for its high x-ray luminosity and has a concentration index
C$\sim$0.49 (\cite{fa91}).

The data for the 10 high-concentration clusters are plotted in Fig.~12
of \cite{dr97}.  There are an average of $\sim$63 cluster galaxies
within a radius of 1450 kpc in these clusters to $M_V$=-20.4.  In
CL1358+62, there are $\sim$114 cluster members within this radius to
$M_V$=-20.4, so CL1358+62 is richer than the average cluster in the
low-$z$ sample.  However, the low-$z$ sample does contain a
high-concentration cluster as rich as CL1358+62: the Coma Cluster.
Furthermore, the density difference between CL1358+62 and the average
low-$z$ reference cluster, 0.3 dex, is comparable to the bin
size in the morphology-density relation plots in \cite{dr97}.

The comparison between the morphological composition of the CL1358+62
sample to $M_V$=-20 and the low-$z$ reference sample is made in Table
3 and Figure 8.  The DF/MF/PVD S0/E ratio, 1.6$\pm$0.3, differs at
only 1.4$\sigma$ confidence from the low-$z$ reference sample ratio of
2.1$\pm$0.2. AD's ratio, 1.1$\pm$0.2, differs from the the low-$z$
reference sample ratio at 3.5$\sigma$ confidence.  

Despite this difference in the S0/E ratio, we stress that both
classifications for CL1358+62 yield fractions of early types (E+S0),
$\sim$72\%, and late types (S), $\sim$28\%, that are identical
within the errors to the low-$z$ reference sample.  We can therefore
draw a robust conclusion from the two sets of morphological
classifications: CL1358+62 does not contain an elevated population of
spiral galaxies compared with low $z$ reference clusters.  This
contrasts with the results of \cite{an97} and and \cite{co98} for
other intermediate $z$ clusters.  Our work suggests that the early
type/spiral classifications are likely to be secure, implying that the
populations of intermediate $z$ clusters vary significantly, even
after accounting for the effects of the low $z$ morphology-density
relation.

This conclusion about the spiral population in CL1358+62 limits the
range of physical models for evolution in CL1358+62, even allowing for
our uncertainty about the E/S0 classifications.  As we mentioned
earlier, the AD classifications have the strong advantage of the same
classifier at low and intermediate $z$.  However, given the different
character of the low-$z$ photographic images and the intermediate-$z$
WFPC2 images, we must acknowledge the possibility of systematic,
redshift-dependent classification uncertainties.  Figure 1 provides
some reason to be cautious about this issue.  Because we do not
understand in detail the reasons for the differences between the two
sets of visual classifications, we cannot be positive that the two sets
of classifications bound our uncertainties.  However, the best we can
do at present is to leave the issue of E/S0 classifications open, and
to explore the consequences of both sets of classifications below.
 
The DF/MF/PvD classifications would imply that cluster evolution from
$z=0.33$ to the present does not affect the cluster morphological
composition or its morphology-density relation within the observed 1.4
Mpc radius aperture.  Figure 9 shows this relation for CL1358+68,
binning the background subtracted data to I=22 ($M_V$=-18.5) radially
about the dominant central galaxy.  Since we are looking only for
radial trends, using the deeper sample is appropriate here.  The
average galaxy density in each of four radial bins (0--1, 1--2, 2--3,
and 3--5 arcmin) is the abscissa for this histogram.  The galaxy
densities in Figure 9 have been normalized to a $M_V$=-20 limit to
allow comparison with the low-$z$ reference sample.  We find that the
morphology-density relation for CL1358+62 is indistinguishable within
the errors from the low-$z$ relation (to $\sim$$M_V$=-20) shown in
Figure 3 of \cite{dr97}.

The AD classifications suggest an evolutionary mechanism that
decreases the fraction of E's, increases the fraction of S0's, while
leaving the fraction of spirals unchanged.  Assuming that there is no
plausible mechanism for directly converting E's into S0's, we can
exclude models that transform the observed $z$=0.33 cluster spirals
into S0's without accretion of additional galaxies.  In order to
transform the $z$=0.33 AD morphological mix into the low-$z$ reference
sample population while accreting the smallest number of galaxies, the
cluster population would increase by 50\%.  Approximately 70\% of the
accreted galaxies would become S0's by the present day, and 30\%
spirals.

\section{The Morphological-Spectral Connection}

While a great deal has been learned about the galaxy population in
intermediate redshift clusters from relatively small samples of
spectra, we must remember that even present-day clusters of galaxies
are a heterogeneous group, differing widely in their degree of
virialization.  Because cluster relaxation may drive galaxy evolution,
the range in cluster galaxy populations may be large at any redshift.
For this reason, it is desirable to connect the morphological and
spectral properties of a large sample of galaxies in each of a number
of clusters directly.

We have spectral classifications from \cite{fi98} sorting each of the
galaxies with spectra into one of the four categories: (1) absorption
lines only, (2) emission lines present, (3) emission lines plus strong
Balmer absorption lines, and (4) k+a (also called E+A).  Category (4)
contains galaxies with the normal absorption lines of E/S0 galaxies
plus strong Balmer absorption lines.  Emission line galaxies have
[OII] 3727 {\AA} emission with equivalent width (EW) $>$5 {\AA}.  If
H$_{\delta}$ absorption of $>$4 {\AA} EW is detected for an emission
line galaxy, the galaxy is classified as emission plus Balmer lines.
Galaxies with [(H$_{\delta}$ + H$_{\gamma}$ + H$_{\beta}$)/3] EW
greater than 4 {\AA}, but [OII] emission with $<$5 {\AA} EW, are
classified as k+a.  We refer the reader to \cite{fi98} for a more
complete summary of the spectral properties of these galaxies and a
comparison with spectra of low-$z$ galaxies.  Table 4 and Figure 10
summarize the comparison between the spectral and morphological
properties for the 191 cluster members in common with \cite{fi98},
using the DF/MF/PvD classifications.

In rough terms, the bulk of the CL1358+62 galaxies follow the
morphological-spectral correlation expected for bright field galaxies
at low redshift: the preponderance of the E and S0 galaxies have pure
absorption line spectra, while the fraction of galaxies with emission
lines rises for the late type spirals.  \cite{dr99} found a similar
behavior for galaxies in the MORPHS sample; see also \cite{pog99}.  We
do know, however, that the percentage of galaxies with emission lines
and strong Balmer absorption lines, $\sim$19\%, is higher than the
$\sim$6\% content of these galaxies in comparable low $z$ clusters
(\cite{dr87}, see also \cite{fi98}).

If we use the AD classifications for the CL1358+62 galaxies, these
conclusions do not change significantly.  The most interesting
difference between the two sets of morphological classifications is
that DF/MF/PvD classify all the E+A galaxies as types S0 to Sb, while
AD classifies these galaxies as having a wider range of morphologies
from E to Sbc.

\section{Discussion and Conclusions}

For CL1358+62, we have acquired a unique data set including a large
mosaic of HST fields and extensive spectroscopy that allows us to
unambiguously determine cluster membership for galaxies with $M_V <
-20$.  We have directly compared the morphological classifications of
two sets of classifiers for the galaxies in CL1358+62.  The two sets
of classifiers agree that (to a limit of $M_V$=-20) the fraction of
early type galaxies (and therefore spirals) in this cluster at
$z$=0.33 is indistinguishable from the fraction in comparable low-$z$
clusters.  In contrast, previous workers, \cite{an97} and \cite{co98},
who also studied WFPC2 images of clusters at z$\sim$0.3, found an
elevated population of spirals compared with low-$z$ reference
samples.  Because our work confirms the reliability of
early-type/spiral classifications from intermediate $z$ WFPC2
observations, we conclude that this is evidence for a dispersion in
the evolution of intermediate-$z$ clusters.

The two groups of classifiers differ on ratio of E to S0 galaxies in
CL1358+62.  DF/MF/PvD find a population of S0 galaxies
(S0/E=1.6$\pm$0.3) that is within 1.4$\sigma$ of the low-$z$ reference
sample, while AD finds a significantly smaller ratio (1.1$\pm$0.2).
This systematic difference is most likely related to the fact that the
transition between S0's and intermediate luminosity E's is rather
gradual.  Many of the intermediate luminosity E's are thought to have
disks, e.g.~\cite{sc98}, \cite{ri90}, and \cite{jo94}.  It may only be
possible to resolve this issue by direct model fitting to images at
low and intermediate $z$.  

Even though we conclude that we have not reliably determined the ratio
of S0's to E's among the early types, our work significantly restricts
possible evolutionary models.  If we accept the MF/DF/PvD
classifications, evolution must preserve the fraction of E's, S0's and
S's as well as the morphology-density relation.  If the AD
classifications are correct, evolution must decrease the fraction of
E's and increase the fraction of S0's while maintaining the fraction
of S's.  A possible mechanism for driving the evolution of the
morphological mix in this latter fashion is accretion of additional
galaxies from the spiral-rich infall region that become predominantly
S0's.  The cluster population within an $\sim$1 Mpc radius must
increase by a minimum of 50\% from $z$=0.5 to the present day in order
convert an intermediate $z$ population rich in E's to a low-$z$
population rich in S0's.  It will be interesting to see whether such
accretion can be produced in simulations of cluster formation.  In
most cluster formation scenarios, massive clusters form by the merging
of pre-existing massive clusters with (presumably) similar populations
of early type galaxies.  It may therefore be difficult to double the
ratio of S0 to E galaxies.

We compare our morphological classifications to our previous spectral
classifications and conclude that the morphologies of the spectrally
``active'' galaxies are as might be expected from the low-$z$ field
population: the galaxies with emission lines are predominently spirals
and the k+a (or E+A) post-starburst galaxies are typically early type
disk galaxies (S0--Sb).

We wish to acknowledge the generous contributions of Alan Dressler to
this paper, including his independent classifications and insightful
comments.  We thank Gus Oemler for checking the classifications and
for thoughtful comments on the manuscript, which led to several
improvements.  We thank Margaret Geller for a critical reading of an
earlier version of the manuscript and helpful comments.  Our referee,
Ian Smail, and our editor, Greg Bothun, both made insightful comments
that helped us clarify the paper.

\appendix

\newpage
\bigskip
\centerline{APPENDIX}

\centerline{SYSTEMATIC EFFECTS OF MEASUREMENT ERRORS}

The effects of classification errors on the distribution of types can
be important. Quantifying these effects is difficult because
morphological classification is a subjective procedure, but we can
gain some insight by considering simple models for the errors.  We
begin by assuming that the numerical type is based on a one
dimensional measurement with a simple, constant error.  This simple
model would imply that any peaks in the distribution of types would be
softened.  If we adopt nominal intrinsic fractions of S:S0:E of
0.20:0.53:0.27, as is approximately correct for the inner 600 kpc of
low redshift clusters (\cite{dr97}), the errors will automatically
decrease the fraction of S0's, and enhance the fraction of E's and
S's.

We can calculate an upper limit to the loss of S0's for our nominal
20:53:27 (S:S0:E) population with this model for the errors.  We
assume that the types are scattered with a mean absolute deviation
(MAD) of 1, and all intermediate types are divided equally between
adjacent types.  The outcome of this experiment is a distribution of
0.27:0.38:0.34 (S:S0:E). In this case, all galaxies scattered beyond
the normal type boundaries were assigned to the boundary type
(i.e.~E).  A MAD of 1 may be an overestimate as the difference in the
types assigned by the two groups of classifiers is of this order. The
intrinsic errors are $\sqrt{2}$ smaller if the errors are
independent. For these smaller errors the resulting distribution would
be 0.24:0.43:0.33 (S:S0:E).  In either case, the results serve to
illustrate that the systematic effects can be very significant.

Let us consider a second, more physical model for the visual
classification errors.  Here, we assume that morphological type is
based on two independent variables with continuous distributions:
bulge-to-total light fraction ($f_b$), and asymmetric features due to
spiral arms ($A$).  This is very similar to the quantitative
classification devised by \cite{ab96}. Galaxies with low $A$ will be
classified as early type ($t < 0$) and then divided into E's or S0's
based on whether $f_b$ is above or below a critical value. It has been
argued that most $L*$ ellipticals have faint disks, e.g.~\cite{ri90}
and \cite{jo94}.  Similarly, the spiral classification will be based
on a combination of $f_b$ and $A$.

If the intrinsic distribution of $f_b$ is flat, then errors in $f_b$
will not change the ratio of S0's to E's.  However, if the intrinsic
distribution of $f_b$ is peaked, the errors will have a systematic
effect.  The sign depends on the details of the intrinsic $f_b$
distributions and the size of the errors.  Contributing to this
uncertainty, the errors can also be asymmetric, if for example, faint
extended disks are missed in noisy data.  The situation is similar for
errors in $A$, where E's and Sa's now share a boundary.  Extensive
simulations are required to estimate the systematic effects of limited
S/N on classification errors, taking into account the the point spread
function of WFPC.  We will undertake this effort in a future paper.

\newpage

\small
\begin{deluxetable}{cccc}
\tablenum{1} 
\tablecaption{ Combination Rules for Morphological Types \label{tab1}}
\tablehead{\colhead{Classifier 1} & \colhead{Classifier 2} 
& \colhead{Classifier 3} & \colhead{Final Classification}}
\startdata
-5 & -5 & -4 & -5\\
-5 & -5 & -2 & -4\\
-5 & -4 & -4 & -4\\
-5 & -4 & -2 & -4\\
-4 & -4 & -2 & -4\\
-2 & -2 & -5 & -4\\
-2 & -2 & -4 & -2\\
-2 & -2 &  0 & -2\\
-2 & -2 &  1 &  0\\
-2 & -2 &  2 &  0\\
-2 &  0 &  1 &  0\\
-2 &  1 &  1 &  0\\
-2 &  1 &  2 &  0\\ 
\enddata
\tablecomments {These rules are used to assign morphological 
types if the three authors did not agree exactly.  If the 
classifications span five numerical types or less between 0 and 10, 
the three numbers are averaged.  The average is rounded up or down 
to the nearest whole numerical type.  If the classifications span 
more than five numerical types, but all are between 1 and 99, we 
assign type 15.  Note that the order of the authors is not
significant.}
\end{deluxetable}

\begin{deluxetable}{ccccccc}
\tablenum{2} 
\tablecaption{Morphological Catalog for CL1358+62 \label{tab2}}
\tablehead{\colhead{Number} & \colhead{X} & \colhead{Y} &
 \colhead{I} & \colhead{cz} & \colhead{Type} & \colhead{Type} \\
       &  &  &  &  & (DF/MF/PvD) & (AD) }
\startdata 
     1 & 2185.23 &  198.20 &        20.85 &  99246 &          -4 &     -4 \\
    12 & 4165.23 &  344.75 &        19.79 &  97702 &          -2 &     -4 \\
    15 & 3424.60 &  366.51 &        19.24 &  95807 &          -2 &      1 \\
    16 & 2155.44 &  387.90 &        19.36 &  97243 &           1 &      1 \\
    25 &  968.84 &  444.59 &        20.54 & 129864 &           6 &      5 \\
    27 & 4576.15 &  456.72 &        19.00 &    -   &           5 &      5 \\
    30 & 1479.89 &  477.46 &        19.38 &  98463 &          -4 &     -5 \\
    33 & 4231.61 &  489.34 &        19.13 & 143869 &           0 &      0 \\
    35 & 4095.85 &  492.36 &        20.47 &  98424 &          -4 &     -3 \\
    37 & 3526.59 &  495.73 &        20.42 &  96743 &          -2 &     -2 \\
    38 & 3377.31 &  502.24 &        18.81 &  99803 &           0 &     -2 \\
    48 & 3302.50 &  558.38 &        20.98 & 164142 &          15 &     99 \\
    51 & 4062.30 &  568.56 &        20.77 &    -   &           8 &      4 \\
    52 & 3024.67 &  571.39 &        16.96 &  48524 &           4 &      3 \\
    53 & 3081.65 &  574.07 &        20.08 &    -   &           4 &      5 \\
    54 & 1525.88 &  580.51 &        15.69 &  25570 &           4 &      5 \\
    61 & 2357.29 &  602.25 &        20.50 & 144702 &          15 &      7 \\
    66 & 4121.06 &  625.81 &        20.99 &    -   &          -2 &     -2 \\
    68 & 3470.13 &  635.49 &        18.95 &  96710 &          -2 &     -2 \\
    70 & 3598.20 &  642.90 &        20.04 &  80512 &           5 &      5 \\
    79 & 4072.28 &  697.05 &        18.92 &  98334 &           0 &     -4 \\
    94 & 2580.14 &  725.66 &        19.64 &  97702 &          -2 &     -2 \\
   101 & 2602.51 &  739.65 &        20.44 & 151970 &           0 &     -5 \\
   103 & 3652.82 &  740.90 &        20.43 &  95699 &           3 &      5 \\
   107 & 3499.88 &  757.63 &        20.99 &    -   &           4 &      5 \\
   114 & 2736.96 &  794.55 &        18.89 &  31012 &           5 &      6 \\
   118 & 2286.07 &  807.47 &        19.08 &  97741 &           0 &     -1 \\
   120 & 1458.64 &  807.67 &        20.02 &    -   &           6 &      5 \\
   125 & 4464.63 &  814.64 &        20.65 &    -   &           5 &      5 \\
   127 & 3372.60 &  818.52 &        19.95 &  99845 &          -4 &     -5 \\
   132 &  907.72 &  840.09 &        20.65 &    -   &          -5 &     -5 \\
   140 & 1319.17 &  854.62 &        19.83 &  98991 &          -4 &     -5 \\
   156 & 1940.02 &  919.07 &        19.74 & 149319 &           1 &     99 \\
   159 & 3641.99 &  927.73 &        20.65 &    -   &           5 &      4 \\
   162 & 2605.02 &  933.83 &        19.94 &    -   &          99 &     -3 \\
   178 &  558.31 & 1001.72 &        20.38 &    -   &         999 &     99 \\
   180 & 1745.41 & 1004.04 &        18.59 &  96677 &           0 &      1 \\
   186 & 2657.55 & 1009.86 &        20.83 &    -   &          -5 &     -5 \\
   192 & 3689.81 & 1034.60 &        20.64 & 100272 &          -2 &     -2 \\
   197 & 2634.39 & 1041.72 &        19.94 & 144632 &           2 &      2 \\
\tablebreak
   205 & 1501.76 & 1068.61 &        19.61 & 120912 &           5 &      1 \\
   206 & 3733.06 & 1075.41 &        20.86 &  98397 &           0 &      0 \\
   222 &  873.00 & 1136.15 &        20.86 &  99237 &          -4 &     -5 \\
   237 & 2777.48 & 1171.76 &        19.64 &  97525 &           5 &      3 \\
   239 & 1674.39 & 1188.96 &        20.92 & 207514 &           7 &     10 \\
   244 & 3455.89 & 1200.72 &        20.08 &    -   &         999 &    999 \\
   248 & 1019.98 & 1207.10 &        20.02 &  99291 &          -2 &     -2 \\
   269 &  698.78 & 1258.32 &        20.54 &    -   &           4 &      6 \\
   277 & 1244.07 & 1269.37 &        20.59 &    -   &           2 &      2 \\
   279 &  999.08 & 1272.41 &        20.44 & 162101 &           4 &      3 \\
   283 & 1400.74 & 1280.19 &        20.93 &  98430 &          -2 &     -2 \\
   285 & 3410.44 & 1288.24 &        20.66 & 100257 &           4 &      4 \\
   286 & 2933.14 & 1291.02 &        20.34 &  98343 &           6 &      4 \\
   295 & 1818.09 & 1320.64 &        17.85 &  25606 &           4 &      3 \\
   299 & 2669.82 & 1334.75 &        18.88 & 100970 &           0 &     -2 \\
   303 & 3046.19 & 1343.52 &        19.71 &  97417 &          -2 &     -2 \\
   304 & 4330.18 & 1348.34 &        17.29 &  23213 &           1 &      1 \\
   314 & 1730.45 & 1364.59 &        20.88 &  97564 &          -2 &     -2 \\
   316 & 1147.10 & 1373.25 &        20.47 &  99893 &           0 &      3 \\
   325 &  946.39 & 1406.99 &        19.25 &  96686 &          -5 &     -5 \\
   326 & 3531.82 & 1407.04 &        20.77 & 100281 &          -2 &      0 \\
   327 & 3089.78 & 1412.64 &        19.03 &  99330 &           1 &      1 \\
   328 & 1340.92 & 1417.67 &        20.94 & 202745 &           0 &      3 \\
   331 & 3058.10 & 1427.59 &        19.08 &  98481 &          -2 &     -2 \\
   333 & 4153.37 & 1437.70 &        18.87 &  98104 &          -5 &     -5 \\
   336 & 3083.01 & 1446.02 &        19.79 &  97627 &          -2 &     -2 \\
   338 & 2994.05 & 1447.54 &        20.62 &    -   &           2 &      1 \\
   352 &  869.48 & 1493.94 &        20.70 &  45442 &           3 &      3 \\
   362 &  715.38 & 1525.08 &        20.21 &  99597 &           1 &      1 \\
   368 & 2960.35 & 1547.05 &        19.92 &  99513 &           0 &      0 \\
   382 &  697.67 & 1576.68 &        20.71 & 149212 &           6 &      5 \\
   385 & 1150.11 & 1586.52 &        18.25 &  98976 &          -4 &     -5 \\
   392 & 3917.25 & 1608.25 &        18.60 &  98541 &           3 &      3 \\
   393 & 2331.21 & 1609.32 &        20.61 &  99237 &          -2 &     -2 \\
   397 & 2423.80 & 1617.24 &        19.02 & 100430 &          -2 &     -2 \\
   399 &  765.00 & 1624.56 &        20.76 & 159692 &           0 &     -2 \\
   415 & 2527.55 & 1680.94 &        19.65 &  98682 &          -4 &     -5 \\
   433 & 2519.00 & 1709.59 &        21.00 &    -   &           6 &      4 \\
   434 & 2712.22 & 1710.19 &        18.44 &  97642 &          -5 &     -5 \\
   435 & 1048.97 & 1710.26 &        20.02 & 190882 &           3 &      2 \\
\tablebreak
   442 & 2657.45 & 1722.45 &        20.88 &  98607 &           1 &     -2 \\
   452 & 3575.83 & 1753.56 &        19.99 &  98086 &           1 &      0 \\
   457 & 2202.26 & 1758.15 &        19.29 &  98688 &          -2 &     -3 \\
   468 & 2524.04 & 1780.65 &        20.30 & 170632 &           2 &      1 \\
   476 & 2164.48 & 1808.28 &        18.87 &  96773 &          -4 &     -5 \\
   484 & 2281.67 & 1835.66 &        17.57 &  97168 &          -5 &     -5 \\
   496 & 4657.77 & 1869.51 &        20.93 &    -   &          -4 &     -5 \\
   499 &  926.31 & 1877.15 &        19.85 & 160364 &          -5 &     -5 \\
   513 & 3385.76 & 1937.13 &        19.49 &  99860 &           0 &     -2 \\
   519 & 2633.79 & 1953.48 &        17.76 &  98691 &          -4 &     -5 \\
   520 & 2267.84 & 1953.77 &        19.98 &  98961 &          -5 &     -5 \\
   521 & 2598.87 & 1954.62 &        20.06 &  99504 &          -2 &     -2 \\
   526 & 1819.56 & 1976.82 &        20.91 &    -   &          -4 &     -5 \\
   529 & 2019.97 & 1981.74 &        20.50 & 118028 &          15 &     -3 \\
   530 & 2604.02 & 1985.11 &        19.63 &  99198 &          -2 &     -2 \\
   533 & 3492.41 & 1992.67 &        19.56 &  97120 &          -5 &     -5 \\
   541 & 2558.48 & 2016.50 &        20.55 &    -   &           0 &      1 \\
   545 & 1341.00 & 2022.51 &        20.10 & 149243 &           0 &     -4 \\
   546 & 4213.68 & 2023.37 &        17.26 &  23768 &           5 &      4 \\
   550 & 2819.78 & 2034.77 &        20.70 & 100970 &           1 &      0 \\
   558 & 2046.76 & 2042.20 &        19.76 &  95753 &          -2 &     -2 \\
   559 & 4563.94 & 2045.51 &        20.58 &    -   &           4 &      4 \\
   561 & 4130.90 & 2050.21 &        19.46 & 207072 &          15 &     99 \\
   562 & 2886.03 & 2061.80 &        18.43 &  69161 &           4 &      3 \\
   563 & 2957.63 & 2065.25 &        20.21 & 100760 &           6 &      0 \\
   564 & 3187.62 & 2070.07 &        19.06 &  97222 &           0 &     -2 \\
   570 & 3067.16 & 2074.98 &        20.66 &  99522 &          -2 &     -5 \\
   581 & 2750.58 & 2107.99 &        19.12 &  98421 &          -2 &     -2 \\
   591 & 2604.10 & 2133.40 &        18.36 &  97762 &          -2 &     -2 \\
   602 & 2350.33 & 2175.38 &        20.38 &  94464 &          -5 &     -5 \\
   606 & 2651.60 & 2182.44 &        19.05 &  96413 &          -2 &     -4 \\
   611 & 1909.12 & 2186.08 &        18.89 &  97612 &          -5 &     -5 \\
   623 & 4308.50 & 2227.50 &        19.93 &  99776 &           3 &      2 \\
   624 & 2123.91 & 2231.85 &        19.46 &  52634 &           0 &     -3 \\
   629 & 5040.18 & 2242.21 &        17.88 &  76299 &           4 &      1 \\
   636 & 2538.86 & 2258.28 &        18.75 &  97342 &          -4 &     -5 \\
   639 & 2777.26 & 2264.35 &        20.40 &  98032 &          -2 &     -2 \\
   641 & 4305.25 & 2279.18 &        20.95 &  23920 &           6 &      4 \\
   643 & 2833.92 & 2289.38 &        20.50 & 148539 &          -2 &     -2 \\
   647 & 2897.12 & 2311.32 &        20.21 &    -   &           6 &      5 \\
\tablebreak
   651 & 3624.13 & 2320.14 &        21.00 & 100652 &          -2 &     -1 \\
   653 & 3096.27 & 2325.88 &        19.87 & 100251 &          -2 &     -4 \\
   655 & 4925.76 & 2334.26 &        20.04 &  98116 &          -2 &     -2 \\
   656 & 3032.45 & 2336.55 &        20.17 &  99708 &          -2 &     -2 \\
   662 & 2762.21 & 2350.96 &        19.15 &  95205 &           1 &      1 \\
   663 & 2002.63 & 2352.95 &        20.24 &  83425 &           0 &     -5 \\
   670 & 4792.86 & 2365.01 &        19.75 &  98125 &          -2 &     -2 \\
   679 & 2891.70 & 2395.08 &        19.51 &  96653 &           0 &     -5 \\
   680 & 4456.98 & 2396.53 &        20.87 & 124924 &           2 &      3 \\
   685 & 4144.73 & 2406.59 &        20.20 &    -   &          -4 &     -5 \\
   698 & 2673.91 & 2444.21 &        20.95 &    -   &           3 &      4 \\
   707 & 3017.95 & 2463.86 &        20.54 &    -   &           0 &     -2 \\
   711 & 2158.68 & 2471.78 &        19.83 &  96536 &          -2 &     -5 \\
   712 & 2420.36 & 2476.45 &        19.69 &  99920 &           0 &      1 \\
   713 & 2213.24 & 2477.51 &        20.55 &    -   &           5 &      4 \\
   717 & 2719.31 & 2486.68 &        18.78 &  96452 &          -2 &     -2 \\
   719 & 2172.06 & 2489.81 &        20.78 &  99339 &          -5 &     -5 \\
   724 & 2281.36 & 2494.96 &        18.85 &  98302 &          -5 &     -5 \\
   730 & 1712.08 & 2511.63 &        19.03 &  99830 &           3 &      1 \\
   732 & 2705.21 & 2518.39 &        20.23 &  96773 &          -2 &     -2 \\
   736 & 3434.65 & 2523.63 &        20.51 &  98451 &           0 &      0 \\
   743 & 2741.21 & 2541.98 &        20.06 &  98272 &          -5 &     -5 \\
   745 & 2878.57 & 2542.91 &        17.95 &  97033 &          -4 &     -5 \\
   748 & 2835.90 & 2554.58 &        20.25 &  97048 &          -5 &     -5 \\
   767 & 2784.63 & 2608.14 &        19.91 &  98302 &          -5 &     -5 \\
   768 & 2566.24 & 2608.94 &        18.44 & 101929 &           1 &      1 \\
   771 & 2851.49 & 2618.35 &        19.79 &  97462 &          -5 &     -2 \\
   774 & 3465.80 & 2625.73 &        18.81 &  97342 &          -2 &     -5 \\
   775 & 2818.67 & 2625.80 &        19.50 &  96824 &          -5 &     -5 \\
   778 & 3368.31 & 2635.15 &        20.05 &  99351 &           1 &      1 \\
   793 & 4179.71 & 2668.46 &        19.19 &  97036 &          -4 &     -2 \\
   794 & 2810.14 & 2668.81 &        19.23 &  98514 &           2 &      1 \\
   799 & 4028.08 & 2680.53 &        19.29 &  96161 &           0 &     -4 \\
   801 & 1766.29 & 2685.75 &        18.63 &  96038 &           1 &      2 \\
   803 & 3671.86 & 2687.29 &        18.34 &  99702 &           1 &      0 \\
   810 & 2801.93 & 2719.27 &        17.11 &  98182 &          -5 &     -5 \\
   813 & 3341.70 & 2731.51 &        19.51 &  96383 &          -2 &     -2 \\
   822 & 4639.03 & 2748.37 &        19.05 &  96997 &           5 &      6 \\
   828 & 2521.07 & 2770.37 &        19.91 &  97912 &           6 &      6 \\
   831 & 2801.07 & 2781.44 &        18.74 &  97672 &          -4 &     -5 \\
\tablebreak
   838 & 2248.16 & 2789.85 &        20.14 &    -   &         999 &     -5 \\
   842 & 2906.30 & 2794.44 &        20.39 & 100670 &         999 &      0 \\
   843 & 4713.68 & 2796.95 &        20.55 &  48385 &           7 &      7 \\
   846 & 2774.56 & 2803.87 &        19.72 &  96563 &          -5 &     -5 \\
   867 & 2612.37 & 2853.73 &        18.37 &  97042 &          -4 &     -2 \\
   872 & 3640.91 & 2868.48 &        20.90 &    -   &          -2 &     -1 \\
   877 & 1235.86 & 2875.59 &        20.67 & 121526 &           1 &      0 \\
   886 & 2138.47 & 2909.08 &        20.20 &  95723 &           8 &      9 \\
   887 & 1185.86 & 2912.30 &        20.47 &    -   &         999 &    999 \\
   888 & 1978.03 & 2915.65 &        19.48 &  96563 &           0 &     -2 \\
   889 & 3904.51 & 2916.67 &        20.36 &  97801 &           0 &      1 \\
   890 & 3514.41 & 2918.22 &        20.68 &  97564 &          -4 &     -2 \\
   898 & 3185.08 & 2942.27 &        20.62 &    -   &          -5 &     -1 \\
   900 & 3332.43 & 2943.17 &        20.63 &  97732 &          -2 &     -2 \\
   905 & 1591.54 & 2966.15 &        20.91 &  98451 &          -5 &     -5 \\
   909 & 3911.90 & 2973.01 &        20.31 & 187822 &           1 &      1 \\
   912 & 1895.44 & 2979.65 &        20.09 &  98751 &          -5 &     -3 \\
   913 & 2748.21 & 2980.39 &        19.26 &  95816 &          -2 &     -2 \\
   922 & 3294.59 & 3021.22 &        19.65 &  98931 &          -2 &     -2 \\
   927 & 2019.29 & 3026.93 &        19.25 &  95274 &          -5 &     -5 \\
   929 & 3951.51 & 3033.18 &        20.10 &  96263 &           0 &     -1 \\
   933 & 4515.25 & 3035.84 &        19.87 &    -   &          -4 &     -5 \\
   938 & 1691.42 & 3049.42 &        19.25 &  71440 &          -4 &      0 \\
   942 & 2829.86 & 3058.61 &        20.59 &  96428 &          -2 &     -5 \\
   943 & 3816.51 & 3058.75 &        19.94 &    -   &          -2 &     -2 \\
   944 & 2420.93 & 3065.24 &        20.94 &    -   &          -5 &     -5 \\
   948 & 4108.78 & 3075.68 &        20.49 &    -   &          -5 &     -4 \\
   961 & 2478.92 & 3095.09 &        19.79 &  99381 &          -5 &     -5 \\
   967 & 3606.49 & 3104.33 &        20.10 &    -   &           4 &      1 \\
   971 & 3131.00 & 3106.51 &        19.94 &  47565 &           5 &      3 \\
   975 & 3024.81 & 3120.83 &        20.95 &    -   &          99 &     99 \\
   978 & 2832.26 & 3123.45 &        20.24 &    -   &          -4 &     -5 \\
   993 & 3323.02 & 3154.25 &        19.05 &  97735 &           0 &     -1 \\
  1003 & 2815.02 & 3176.78 &        20.76 & 100763 &          -5 &     -5 \\
  1012 & 2910.90 & 3204.76 &        20.77 &    -   &          -2 &     -4 \\
  1019 & 1598.82 & 3220.61 &        20.02 &  98718 &           0 &      1 \\
  1020 & 2661.84 & 3222.75 &        18.46 &  83578 &           3 &      1 \\
  1027 & 2954.26 & 3244.74 &        20.28 &  96686 &           0 &     -2 \\
  1036 & 4346.84 & 3277.98 &        19.22 &  99165 &          -2 &     -2 \\
  1041 & 4453.92 & 3289.61 &        19.78 &  98649 &          -2 &     -2 \\
\tablebreak
  1042 & 3771.63 & 3293.33 &        20.93 &    -   &          -2 &      1 \\
  1045 & 4485.09 & 3295.86 &        19.95 &  44686 &          99 &     99 \\
  1049 & 2469.20 & 3302.45 &        20.58 &    -   &           0 &      0 \\
  1062 & 3268.23 & 3338.69 &        18.84 &  98182 &           0 &     -1 \\
  1064 & 4749.89 & 3342.41 &        19.45 &  96182 &           5 &      2 \\
  1069 & 1366.71 & 3359.61 &        19.95 &  99060 &           0 &     -2 \\
  1075 & 4869.37 & 3367.38 &        19.90 &  62244 &           1 &     -2 \\
  1080 & 2041.26 & 3397.59 &        20.86 &  98421 &          -5 &     -5 \\
  1091 & 3315.98 & 3426.51 &        20.51 &  99579 &          -2 &      0 \\
  1100 & 4416.75 & 3462.54 &        20.73 & 159095 &          15 &     99 \\
  1105 & 2881.63 & 3475.46 &        18.79 &  98323 &          -2 &     -2 \\
  1108 & 2836.73 & 3485.88 &        19.42 &  99081 &           1 &     -1 \\
  1121 & 1960.44 & 3512.66 &        20.68 &    -   &          -4 &     -5 \\
  1127 & 2942.66 & 3523.15 &        20.76 & 159362 &           5 &      3 \\
  1141 & 4517.75 & 3545.75 &        19.93 &  99657 &          -2 &     -2 \\
  1146 & 4117.26 & 3561.40 &        18.27 & 100080 &          -5 &     -5 \\
  1153 & 2566.37 & 3583.80 &        20.14 &  61935 &           9 &     99 \\
  1157 & 2487.08 & 3591.79 &        18.64 &  62207 &           9 &      5 \\
  1158 & 3279.02 & 3592.19 &        20.16 &  99935 &          -2 &     -2 \\
  1161 & 2923.10 & 3597.44 &        20.01 &    -   &          -5 &     -5 \\
  1163 & 1348.04 & 3601.03 &        20.12 & 144364 &           1 &      3 \\
  1177 & 2091.53 & 3633.12 &        20.39 & 216269 &          15 &      1 \\
  1178 & 2361.82 & 3635.18 &        20.30 &  99234 &          -2 &     -1 \\
  1192 & 3212.40 & 3665.38 &        20.39 &  97375 &          -5 &     -5 \\
  1200 & 2024.55 & 3679.35 &        20.59 &    -   &           5 &      7 \\
  1207 & 3978.09 & 3693.67 &        20.68 & 100221 &          -4 &     -2 \\
  1209 & 3395.58 & 3703.43 &        19.10 &  83882 &           1 &      1 \\
  1211 & 2923.22 & 3706.89 &        20.79 &  99177 &          -4 &     -5 \\
  1214 & 3909.03 & 3714.53 &        19.51 & 100323 &           1 &      1 \\
  1223 & 2310.92 & 3752.63 &        20.49 &    -   &          -5 &     -5 \\
  1231 & 3481.26 & 3785.53 &        20.32 &  99018 &          -2 &      0 \\
  1235 & 2884.78 & 3791.48 &        20.67 &    -   &          -4 &     -2 \\
  1239 & 2759.52 & 3812.84 &        20.00 &  62264 &           3 &      2 \\
  1240 & 2878.69 & 3813.83 &        20.24 & 122164 &           4 &      7 \\
  1253 &  995.43 & 3851.25 &        19.37 &  97093 &           2 &      4 \\
  1255 & 2908.32 & 3863.10 &        20.54 &    -   &           0 &      3 \\
  1258 & 2180.86 & 3869.70 &        20.60 &    -   &          -5 &     -5 \\
  1265 & 3358.24 & 3888.19 &        19.60 &  99896 &          -2 &     -2 \\
  1272 & 3738.88 & 3907.25 &        20.85 &  99501 &          -2 &     -1 \\
  1273 & 2182.60 & 3907.71 &        20.83 &    -   &          -5 &      0 \\
\tablebreak
  1275 & 2223.31 & 3912.98 &        20.08 &  99141 &          -2 &     -1 \\
  1285 & 1476.03 & 3937.66 &        19.84 & 188687 &           1 &      1 \\
  1287 & 2727.83 & 3947.24 &        20.20 &    -   &          -2 &     -2 \\
  1297 & 2476.43 & 3965.12 &        20.88 &  97102 &          -2 &     -2 \\
  1298 & 2291.46 & 3965.96 &        20.79 &    -   &          -2 &     -2 \\
  1299 & 1553.40 & 3967.72 &        20.25 &  98964 &           3 &      3 \\
  1301 & 1428.04 & 3978.07 &        19.40 &  99342 &           0 &     -2 \\
  1303 & 3180.35 & 3982.62 &        18.98 &  96901 &           0 &     -2 \\
  1308 & 1799.52 & 3992.55 &        20.30 &    -   &           5 &      3 \\
  1322 & 1469.20 & 4032.65 &        20.90 &    -   &          -2 &     -2 \\
  1330 & 2626.63 & 4050.56 &        19.91 &  97051 &          15 &     10 \\
  1335 & 3007.32 & 4063.99 &        20.39 &    -   &          -4 &     -4 \\
  1343 & 2119.25 & 4082.59 &        17.77 &  99156 &          -5 &     -5 \\
  1357 & 1560.05 & 4129.23 &        19.78 &  97000 &          -5 &     -2 \\
  1367 & 2972.68 & 4167.09 &        18.02 &  98299 &          -5 &     -5 \\
  1370 & 3118.58 & 4169.24 &        19.71 &  97738 &          -2 &     -5 \\
  1372 & 3911.67 & 4178.76 &        20.50 &  98472 &          -5 &     -5 \\
  1374 & 3888.45 & 4183.67 &        20.52 &  97699 &          -2 &     -2 \\
  1376 & 3489.91 & 4187.61 &        19.91 &  98275 &          -2 &     -2 \\
  1377 & 2964.53 & 4192.76 &        19.23 &  97975 &          -5 &     -5 \\
  1379 & 3412.68 & 4193.30 &        20.93 &    -   &           0 &      2 \\
  1382 & 2988.76 & 4209.19 &        20.23 &    -   &           3 &      4 \\
  1387 & 2761.87 & 4222.14 &        19.18 &  99417 &          -5 &     -5 \\
  1395 &  848.83 & 4239.90 &        19.39 &  99771 &          -2 &      1 \\
  1405 & 3807.69 & 4288.35 &        20.12 &  99285 &          -4 &     -5 \\
  1409 & 3951.25 & 4295.39 &        19.69 &  54964 &          -2 &     -2 \\
  1411 & 2500.75 & 4322.96 &        19.74 &    -   &         999 &     -2 \\
  1413 & 3281.08 & 4328.24 &        19.37 &  99774 &           2 &      2 \\
  1414 & 1303.46 & 4329.27 &        20.65 &  95615 &           5 &      3 \\
  1415 &  288.53 & 4331.63 &        18.70 &  98844 &          -5 &     -5 \\
  1423 & 3633.52 & 4361.01 &        20.56 &  98799 &          -2 &     -2 \\
  1430 & 4193.88 & 4380.15 &        20.61 &  99222 &         999 &      3 \\
  1431 & 3125.60 & 4381.56 &        20.86 &    -   &          -2 &     -4 \\
  1438 &  574.34 & 4410.17 &        18.72 &  81105 &           2 &      1 \\
  1439 & 3020.74 & 4422.23 &        19.27 &  99339 &           0 &     -5 \\
  1447 &  755.69 & 4452.72 &        20.97 & 171424 &           3 &      4 \\
  1454 & 3375.34 & 4469.90 &        20.28 & 100859 &          -2 &     -3 \\
  1461 & 4207.84 & 4489.60 &        20.98 &  97279 &          -2 &     -5 \\
  1466 & 3188.31 & 4501.44 &        20.62 &    -   &           5 &      5 \\
  1481 &  810.82 & 4540.51 &        20.61 &  99791 &          -2 &      2 \\
\tablebreak
  1483 &  825.24 & 4542.01 &        19.36 &  99791 &          -5 &     -5 \\
  1488 &  661.75 & 4553.49 &        19.55 &  80904 &          -2 &     -4 \\
  1524 &  943.52 & 4664.34 &        20.89 & 190098 &           6 &      5 \\
  1529 & 3450.04 & 4681.31 &        20.81 & 114632 &          15 &     10 \\
  1540 & 3674.58 & 4731.74 &        18.01 &  21317 &           5 &      6 \\
  1541 & 2265.25 & 4732.11 &        20.43 &  98778 &          -5 &     -5 \\
  1559 & 3242.72 & 4783.73 &        20.72 & 100380 &           1 &      1 \\
  1563 & 4731.77 & 4795.22 &        20.88 &    -   &          -2 &     -1 \\
  1567 &  768.30 & 4815.43 &        20.88 & 100284 &          99 &     10 \\
  1568 & 4014.76 & 4819.25 &        19.46 &  36222 &          99 &     -5 \\
  1571 & 2338.11 & 4825.66 &        19.97 &  83803 &           4 &      5 \\
  1588 & 4185.04 & 4870.66 &        20.95 & 126167 &           0 &     -4 \\
  1594 & 1852.07 & 4887.76 &        19.87 &    -   &         999 &    999 \\
  1621 & 1467.74 & 4980.41 &        20.03 &    -   &          -5 &    999 \\
  1623 & 1937.90 & 4990.12 &        20.86 & 187086 &          -2 &     -2 \\
  1634 & 3158.57 & 5037.47 &        19.28 &  31985 &           9 &      5 \\
  1664 & 2256.63 & 5240.59 &        20.61 & 248908 &           5 &      2 \\
  1666 & 2243.04 & 5270.84 &        19.15 &  99552 &           1 &      2 \\
\enddata
\tablecomments {We consider those galaxies with 94318$<$cz$<$102532 to
be cluster members.  Each pixel for the X and Y positions corresponds
to 0.1$^{\prime\prime}$, and the cluster center is taken to be
coincident with the center of galaxy \#810 at 2801.93, 2719.27.
This corresponds to R.A. [1950] 13$^h$58$^m$20$^s$.7, decl.~[1950]
62$^{\circ}$45$^{\prime}$33$^{\prime\prime}$.}
\end{deluxetable}

\begin{deluxetable}{cccccc}
\tablenum{3} 
\tablecaption{Evolution of the Morphological Mix in CL1358+62 (\%) 
\label{tab3}}
\tablehead{\colhead{Type} & \colhead{CL1358} & \colhead{CL1358} &
 \colhead{Low Z}\\
   & (DF/MF/PvD) &   (AD)      &   (AD)   }
\startdata 
E  & 27$\pm$4 & 35$\pm$5  & 23$\pm$2 \\
S0 & 44$\pm$6 & 38$\pm$5  & 49$\pm$3 \\
S  & 29$\pm$5 & 27$\pm$4  & 27$\pm$2 \\
\enddata
\tablecomments {We assume that the unclassified galaxies in the
CL1358+62 sample (1.5\%) have the same morphological mix as the
classified galaxies. The classifications for CL1358 in column 2
are the combined work of DF, MF and PvD (see text for
details).  The classifications in column 3  are the
independent work of AD. The limiting magnitude is $\sim$$M_V$=-20, and
the limiting radius is $\sim$1400 kpc.}
\end{deluxetable}

\begin{deluxetable}{ccccc}
\tablenum{4} 
\tablecaption{Spectral vs.~Morphological Properties of Galaxies in CL1358+62
\label{tab4}}
\tablehead{\colhead{Type} & \colhead{Pure Absorption} &
\colhead{Emission Lines} & \colhead{Emi.~+ Balmer Abs.} & 
\colhead{k+a}} 
\startdata 
E      & 46   &  1.5 &  0.5 &  0\\
S0     & 76.5 &  2.5 &  1.5 &  7\\
Sa-Sb  & 26.5 &  5   &  2   &  4\\
Sbc-Irr&  3 &  4 &  6 &  0\\
Merger &  0 &  1 &  1 &  0\\
?      &  3 &  0 &  0 &  0\\
\enddata
\tablecomments {The DF/MF/PvD morphological classifications 
are used here; the spectral classifications are from \cite{fi98}. 
Ambiguous types (E or S0, S0 or Sa) are split equally between the two
adjacent types, accounting for the fractional galaxies.}
\end{deluxetable}

\begin{figure}
\epsscale{0.8}
\plotone{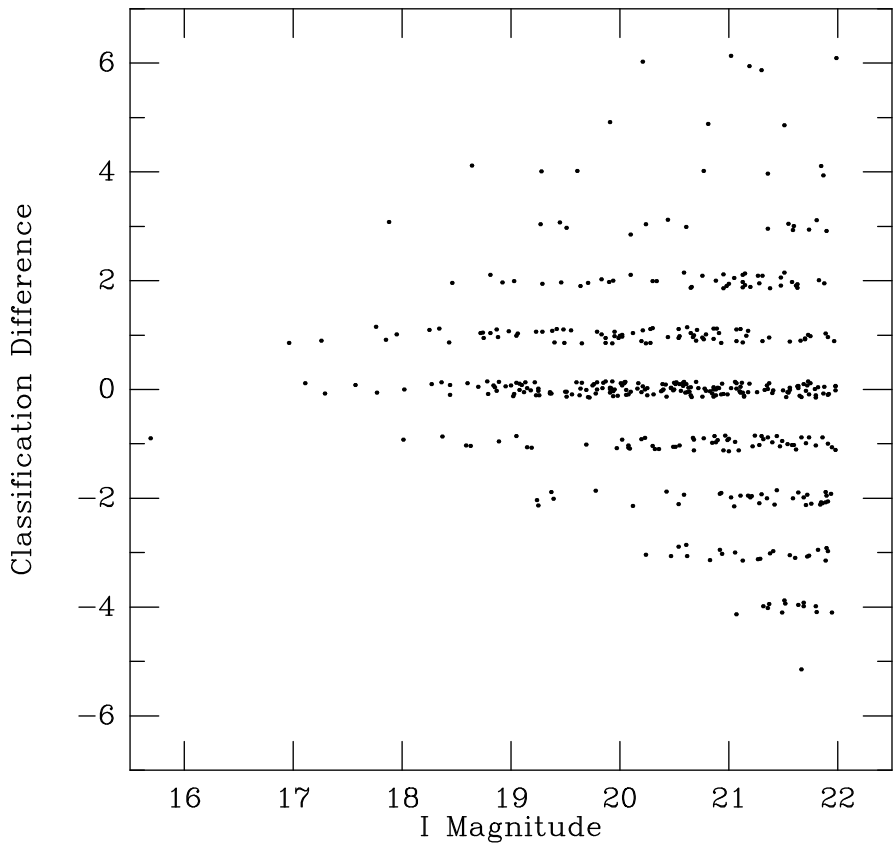}
\caption{The difference between the DF/MF/PvD
classifications and the AD classifications as a function of $I$
magnitude.  For the purposes of Figures 1--3 the numbers corresponding
to morphological type have been shifted and condensed.  For these
figures, E is -3, E/S0 or S0/E is -2, S0 is -1, and S0/Sa or Sa/S0 is
0.  The later morphological types have been assigned the conventional
numbers described in the text.  The difference between the
classifications rises steeply below $I=22$.}
\end{figure}
\clearpage

\begin{figure}
\epsscale{0.8}
\plotone{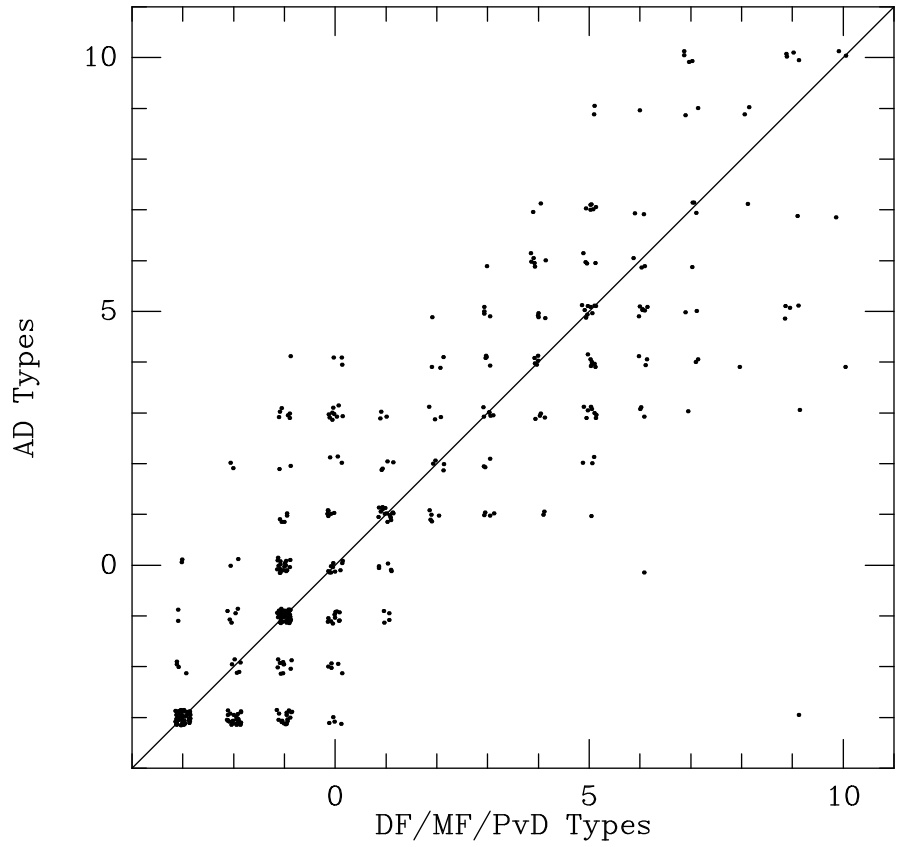}
\caption{A scatter diagram comparing the two sets of
classifications.  See the text or the caption of Figure 1 for the
conversion to morphological types.}
\end{figure}
\clearpage

\begin{figure}
\epsscale{0.8}
\plotone{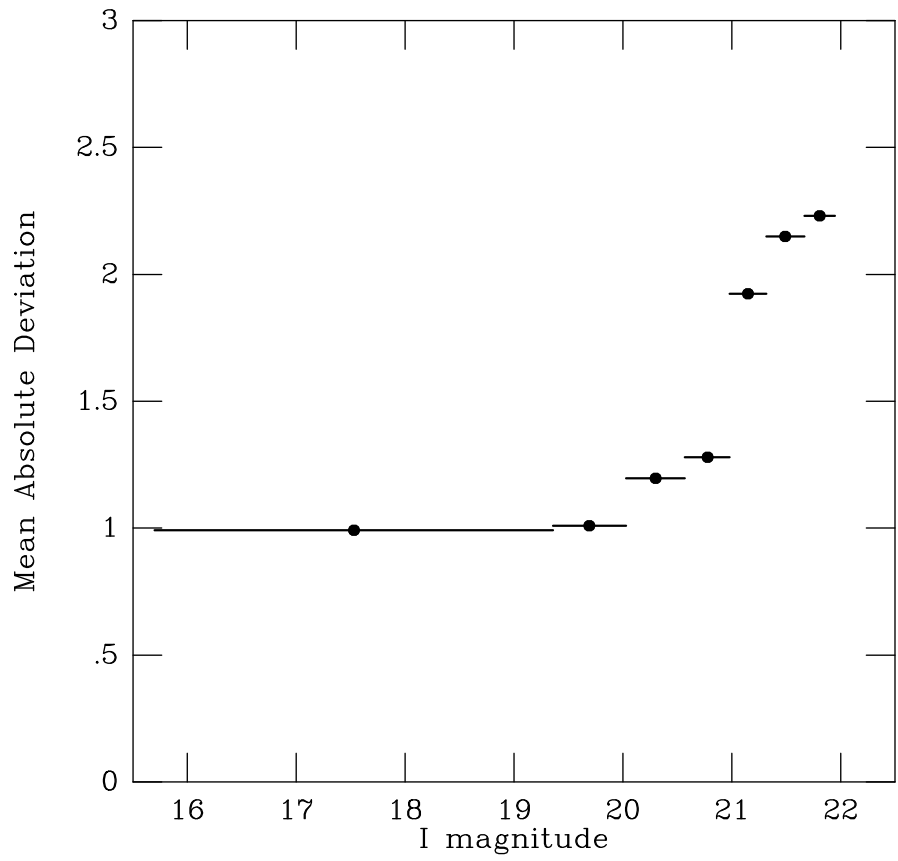}
\caption{The mean absolute deviation between the two sets
of classifications as a function of $I$ magnitude, normalized to the
mean absolute deviation of a Gaussian function with a standard
deviation of 1.}
\end{figure}
\clearpage

\begin{figure}
\epsscale{0.8}
\plotone{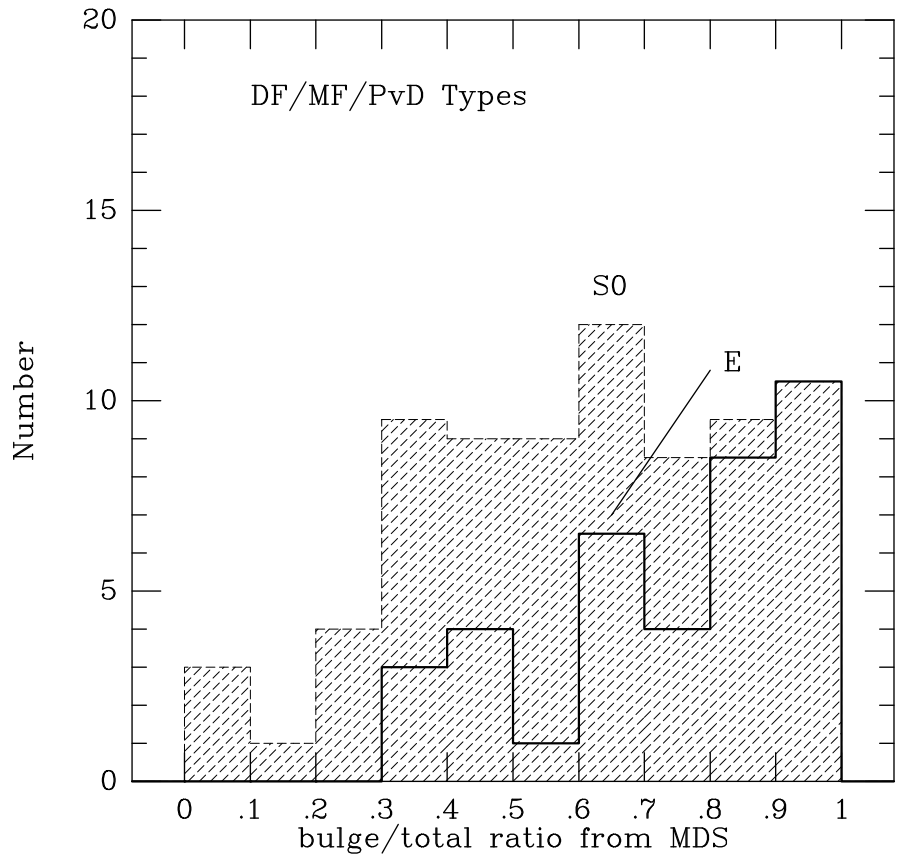}
\caption{A comparison of the MDS bulge to total light
ratios for galaxies that DF/MF/PvD classify as S0 and E, excluding
ambiguous types -4 and 0.  The magnitude limit is $I=21$.  MDS
structural parameters are available for $\sim$70\% of the CL1358+62
galaxies.}

\end{figure}
\clearpage

\begin{figure}
\epsscale{0.8}
\plotone{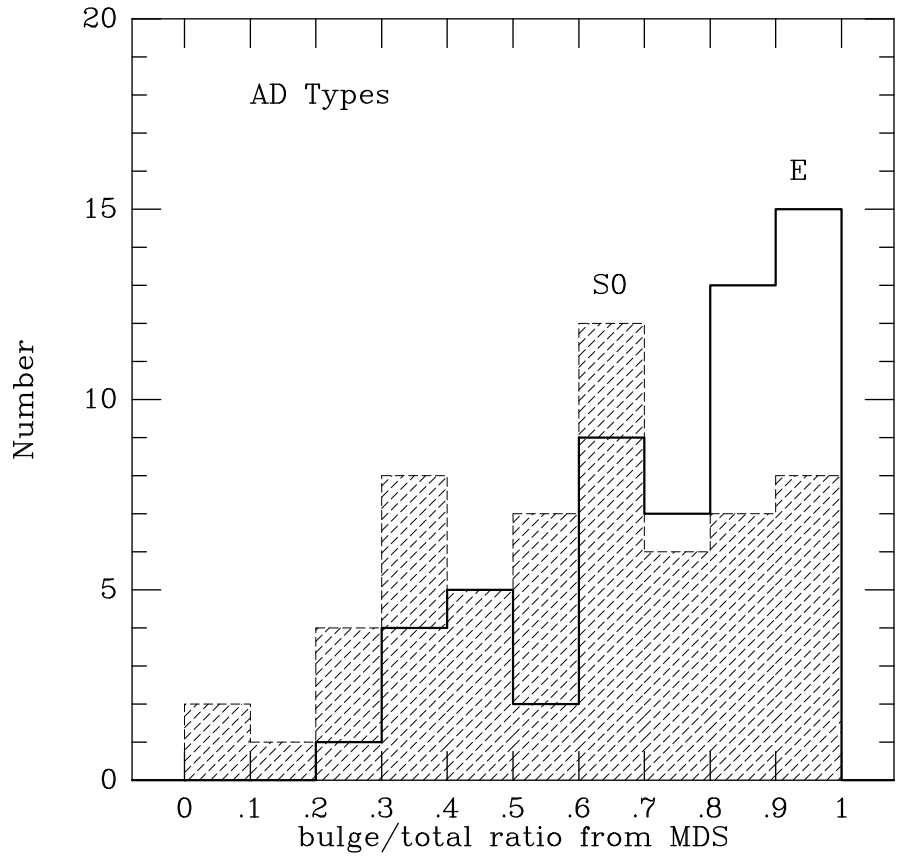}
\caption{A comparison of the MDS bulge to total light
ratios for galaxies with $I<21$ classified by AD as S0 and E.}
\end{figure}
\clearpage

\begin{figure}
\epsscale{0.8}
\plotone{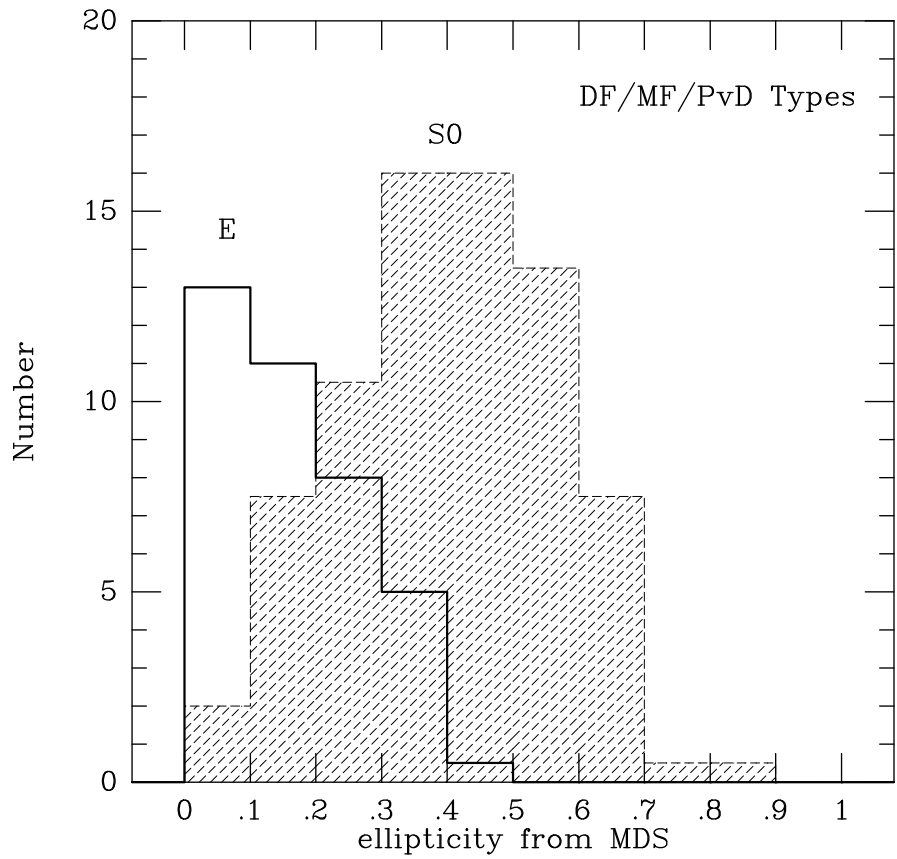}
\caption{The ellipticity distributions for the galaxies
with $I<21$ classified by DF/MF/PvD as S0 and E. Ambiguous types -4
and 0 were excluded.} 
\end{figure}
\clearpage

\begin{figure}
\epsscale{0.8}
\plotone{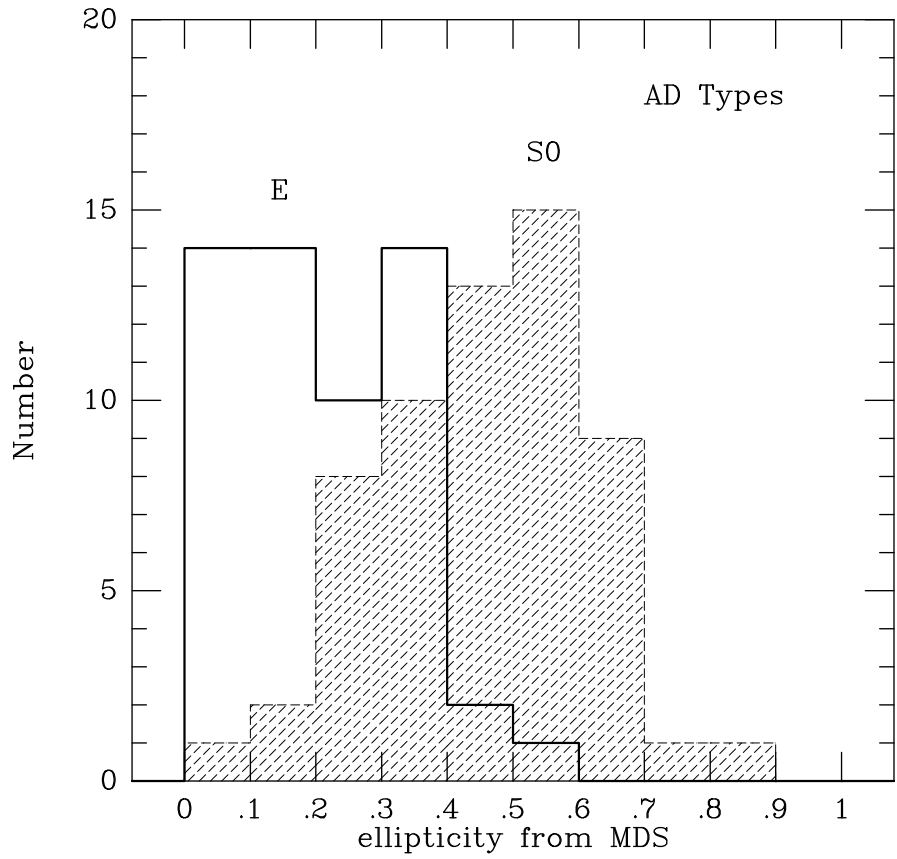}
\caption{The ellipticity distributions for the galaxies
with $I<21$ classified by AD as S0 and E.} 
\end{figure}
\clearpage

\begin{figure}
\epsscale{0.8}
\plotone{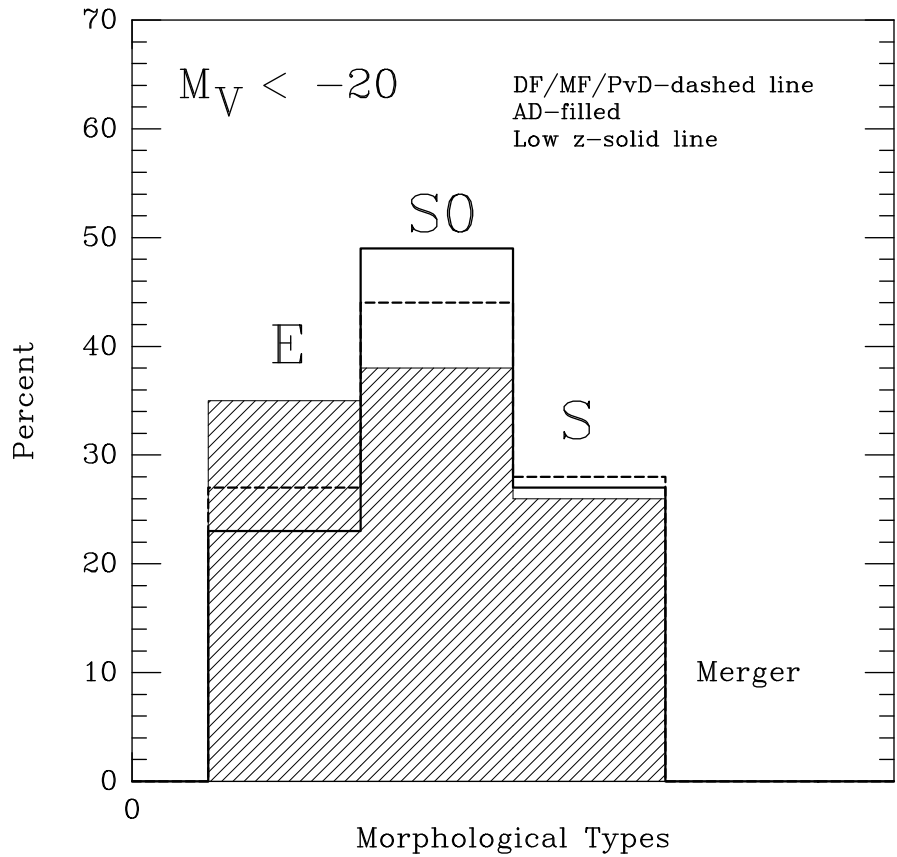}
\caption{The morphological composition of CL1358+62 compared with the low-$z$
reference sample of \cite{dr97}.  In both cases, the limiting magnitude
is $M_V$$\sim$-20 and the limiting radius is $\sim$1400 kpc.  Both
sets of classifications for CL1358+62 are shown.}

\end{figure}
\clearpage

\begin{figure}
\epsscale{0.8}
\plotone{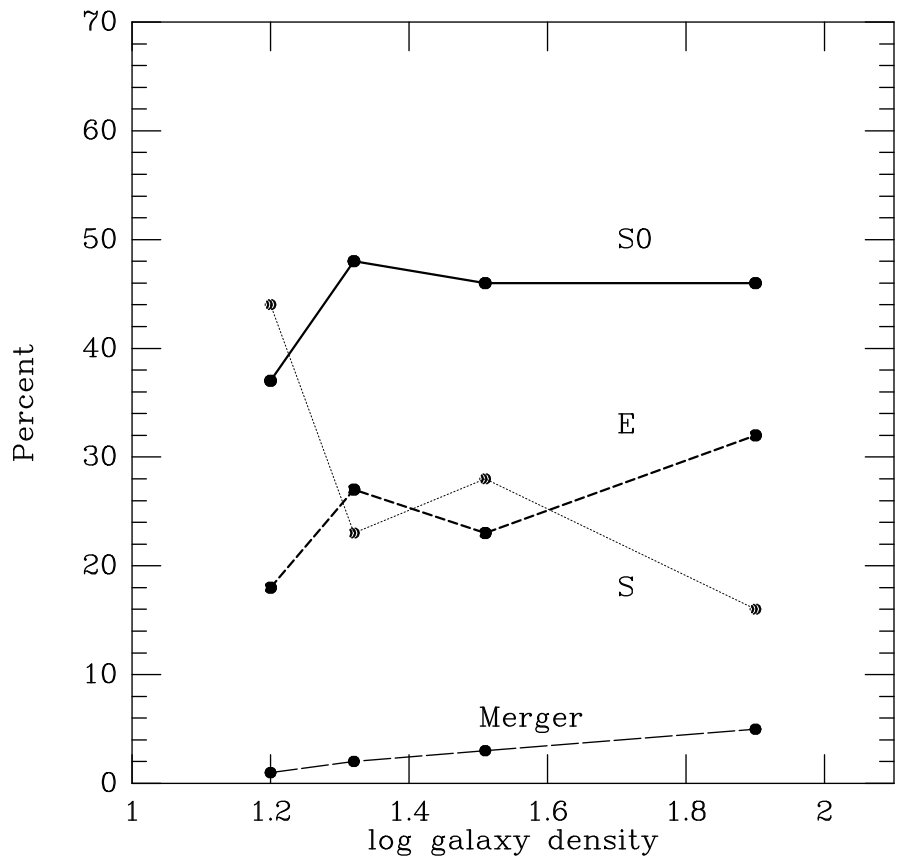}
\caption{The morphology-density relation in CL1358+62 using the DF/MF/PvD
classifications.  This relation is almost identical to that plotted in
Figure 12 of \cite{dr97} for the low-$z$ reference sample.}
\end{figure}
\clearpage

\begin{figure}
\epsscale{0.8}
\plotone{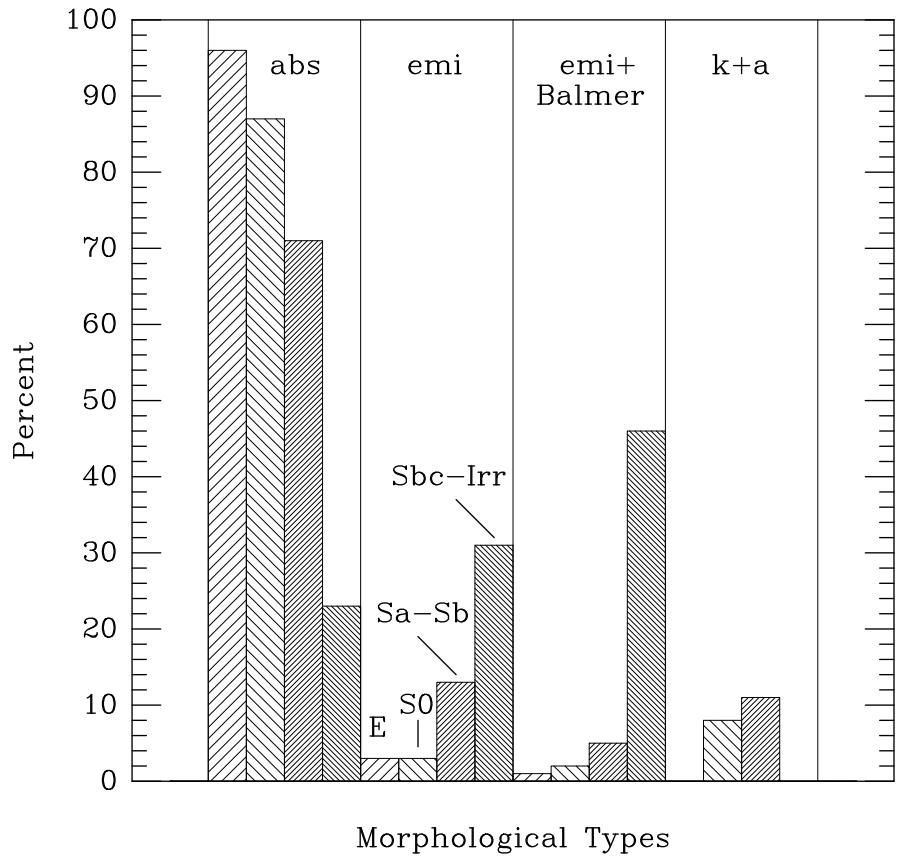}
\caption{The relation between the morphological and spectral properties of the
CL1358+62 galaxies, using the DF/MF/PvD classifications.}
\end{figure}
\clearpage

\end{document}